\documentclass[aps,pra,reprint,groupedaddress]{revtex4-1}
\usepackage{amsmath,graphicx,hyperref}
\usepackage[FIGTOPCAP,raggedright,nooneline,bf,footnotesize]{subfigure}
\usepackage{indentfirst}
\usepackage[usenames,dvipsnames]{pstricks}
\usepackage{overpic}
\usepackage{xcolor}
\usepackage{pict2e}

\newcommand{\figref}[1]{Fig.~\ref{#1}}
\newcommand{\tabref}[1]{Tab.~\ref{#1}}
\renewcommand{\eqref}[1]{Eq.~(\ref{#1})}

\begin{document}
\title{Remote temporal wavepacket narrowing}


\author{Karolina Sedziak}
\author{Miko\l aj Lasota}
\author{Piotr Kolenderski}

\email{kolenderski@fizyka.umk.pl}
\affiliation{Faculty of Physics, Astronomy and Informatics, Nicolaus Copernicus University, Grudziadzka 5, 87-100 Toru\'{n}, Poland} 

\date{\today}

\begin{abstract}
	Quantum communication and clock synchronization protocols can be significantly enhanced by the careful preparation of the wavepackets of the produced photons.  Following the theoretical proposal published in [Optica, \textbf{4}, 84 (2017)], we experimentally demonstrate the effect of a remote temporal wavepacket  narrowing of a heralded single photon. It is performed by utilizing the time-resolved measurement on the heralding photon which is frequency entangled with the heralded one. We also investigate optimal photon pair source characteristics that minimize the heralded wavepacket width. 
\end{abstract}

	\maketitle
	
\section{Introduction}

The phenomenon of entanglement can be certainly seen as the essence of quantum theory. It leads to a plethora of counterintuitive effects, allowing for substantial improvement of numerous applications. For example, polarization entanglement is an important resource for protocols of secure information transfer \cite{Ekert1991,Gisin2002,Gisin2007}, while spatial entanglement is essential for ghost imaging \cite{Pittman1995, Shih2008} and spatial quantum states encoding \cite{SolisProsser2011,Solis-Prosser2013,Varga2014}. On the other hand, spectral entanglement can be utilized to improve the quantum communication (QC) security \cite{Gisin2007, Sedziak2017}.

Photonic implementations of QC protocols suffer from many setup imperfections, plaguing realistic single-photon sources, communication channels and detection systems. All of them together limit the maximal secure distance of information transfer. One of the most effective ways to reduce this limitation is to use the temporal filtering method to decrease the amount of noise registered by single-photon detectors \cite{Patel12}. However, full potential of this method cannot be used if the single-photon wavepacket is affected by the temporal broadening effect. This turns out to be a common problem in fiber-based QC systems, where the signals propagate through dispersive media. Recently, a method to significantly reduce the problem of temporal broadening was proposed \cite{Sedziak2017}. It is based on an appropriate preparation of spectrally correlated photon pairs \cite{Kim2002,Kim2005a,Lutz2013,Lutz2014}.

Remote quantum clock synchronization is another application benefiting from correlated pairs of photons. The idea of utilizing a number of frequency-entangled wavepackets for the enhancement of its timing accuracy was proposed in Ref.~\cite{Giovannetti2001a}. The framework discussed by the authors of this article assumed that the photon pair source was pumped with continuous wave (CW) laser. This means that the spectra of the emitted photons are negatively correlated. An experimental realization of the remote quantum clock synchronization protocol using CW laser was reported by Valencia \emph{et al.} in Ref.~\cite{Valencia2004}. The main idea of this work relies on measuring the statistics of relative arrival times of pairs of correlated photons after their propagation on certain distances. The precision of synchronization depends on the number of detected pairs, temporal widths of their wavepackets, and the quality of the best fit of the assumed theoretical model. More recently, in Ref.~\cite{Quan2016},  Quan \emph{et al.} demonstrated another synchronization method utilizing additional feedback mechanism based on Hong-Ou-Mandel interference, allowing to compensate temperature-related fluctuations of the channel length. However, all those quantum clock synchronization procedures are limited to the time scales below the characteristic fluctuation time. Another major limitation for them is the temporal broadening of the photon pair wavepacket during their propagation in telecommunication fibers.

In this work, we experimentally investigate the problem of remote preparation of a single-photon wavepacket by a spontaneous parametric down-conversion (SPDC) source. Following the theoretical proposal published in Ref.~\cite{Sedziak2017}, we show how adjusting the spectral entanglement and applying time-resolved heralding procedure can substantially narrow the wavepacket of the propagated photons in comparison with the classical case. We also discuss the problem of optimizing the SPDC source for applications utilizing telecommunication fibers of given length and dispersion coefficient.  We discuss our results in the context of improving the performance of remote quantum clock synchronization schemes.

%


%

\section{Wavepacket narrowing}

\begin{figure}[ht!]
	\centering
	\includegraphics[width=0.95\columnwidth]{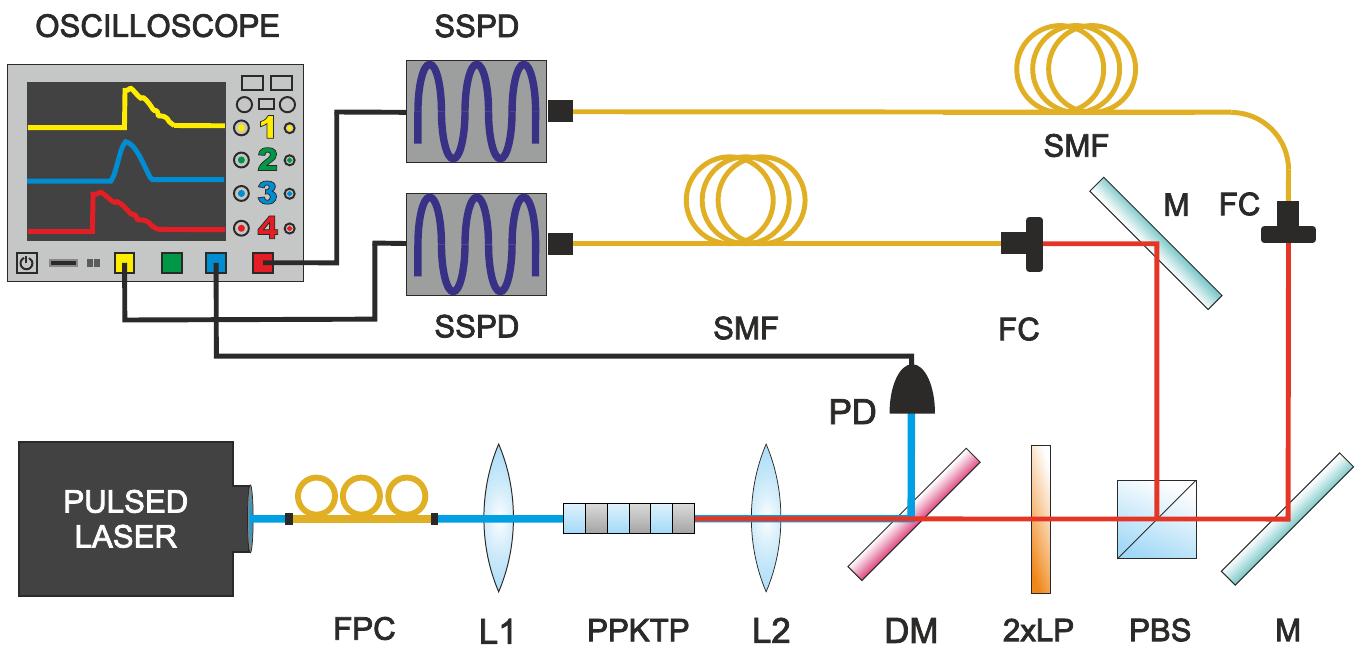} 
	\caption{The experimental setup. Pulses from the femtosecond Ti:Sapphire laser are coupled to a fiber. Their  polarization is set by a fiber polarization controller FPC. The pump is focused using the lens  L1  (plano-convex F=$125$ mm) in the type II PPKTP nonlinear crystal. The resulting photon pairs are collimated using the lens L2 (plano-convex , F=$70$ mm). A dichroic mirror DM (Semrock FF875-Di01) separates the unconverted laser beam and the SPDC photons. Two longpass filters LP  (Semrock BLP01-1319R-25) clean the remainder of the pump. Next, polarizing beam splitter PBS separates the SPDC photons with respect to their polarization. They are subsequently  coupled to single mode fibers SMF (Corning SMF28e+) using fiber collimators FC (F=8.0 mm) and mirrors M. The oscilloscope monitores the outputs of superconducting nanowire single-photon detectors SSPD and the fast photodiode PD.}
	\label{fig:setup}
\end{figure}

The experimental setup utilized to measure the arrival time distribution of the photon pairs is depicted in \figref{fig:setup}. The parametric down conversion process takes place in  a type-II $10$ mm-long PPKTP with a poling period of $46.2$ $\mu $m, designed for collinear phase matching $780$ nm $\rightarrow 2 \times 1560$ nm. 
The crystal was pumped by a pulsed Ti:Sapphire laser coupled to a single-mode fiber, with the central wavelength $780.1(1)$ nm and the repetition rate $80.14(2)$ MHz. A dichroic mirror was used to separate the pump beam and a photon pair. The photodiode illuminated with the pump beam delivered the reference signal, carrying the information on the emission time of every pair. Subsequently the orthogonally polarized SPDC photons were separated by a polarizing beam splitter and coupled into two $10$ km-long  standard single-mode fibers (SMFs). The measured  value of the group velocity dispersion was $2\beta=-2.27\times10^{-26}\frac{\mathrm{s}^2}{\mathrm{m}}$ at $1560$ nm, which corresponds to 17.6 ps/(nm$\cdot$km). The longpass filters were used to clean the remaining pump beam. The photons were detected using a superconducting nanowire detectors (SSPDs). An oscilloscope measured the electric pulses from the SSPDs and the photodiode, and recorded the respective waveforms, which were post-processed in order to obtain the time stamps for the photon detection. The triggering system was set to record only coincidences from the two SSPD channels in a given time window. The resulting data set was composed of pairs of photon detection times referenced to the laser pulse. The detection system and electronics contributed to the timing jitter, which standard deviation was of the order of $45$ ps for the SSPDs and around $4$ ps for the photodiode. 
\begin{table}[h]
	\centering
	\begin{tabular}{c|c|c|c}
		Data set & \#$1$ & \#$2$ & \#$3$ \\ \hline		
		$\Delta\lambda$ & $12.47(8)$ nm & $0.929(12)$ nm & $0.551(14)$ nm\\ \hline
		$\tau_p$        & $71.82(42)$ fs  & $0.964(13)$ ps  & $1.62(4)$ ps \\ \hline
		$\rho_t$          & $0.9551(2)$ & $-0.1483(14)$ & $-0.4443(11)$ \\ \hline
		$\tau_1$        & $1.136(2)$ ns & $0.23607(24)$ ns & $0.2146(2)$ ns \\ \hline
		$\tau_2$        & $1.312(2)$ ns & $0.25285(25)$ ns & $0.23130(21)$ ns\\ \hline
		$\tau_{1h}/\tau_{1}$        & $29.49(41)\%$ & $96.9(7)\%$ & $87.9(5)\%$
	\end{tabular}
	\caption{The three pump bandwidths, $\Delta\lambda$, utilized in the experiment (FWHM), the corresponding pulse duration, $\tau_p$, the best fit parameters $\rho$, $\tau_1$, $\tau_2$ for the statistics of arrival times of SPDC photons to the detectors, and the ratio of reduced temporal width of the heralded photon to the respective temporal width of non-heralded photon, $\tau_{1h}/\tau_1$.}
	\label{tab:fitvalues}
\end{table}

The spectral wavefunction of a fiber-coupled pair of photons generated in SPDC process can be parametrized using the quantities directly describing the properties of the biphoton state, namely its spectral correlation coefficient $\rho$ and spectral widths $\sigma_1$, $\sigma_2$. Within this parametrization the probability of  detecting two photons at time $t_1$ and $t_2$ is given by a bivariate normal distribution:
\begin{multline}
p(t_1,t_2)=\frac{1}{\sqrt{\pi } \sqrt{\tau_1\tau_2 \sqrt{1-\rho_t ^2}}}\\\times\exp \left(-\frac{1}{2 \left(1-\rho_t ^2\right)}\left(\frac{t_1^2}{\tau_1^2}+\frac{t_2^2}{\tau_2^2}-\frac{2 t_1 t_2 \rho_t }{\tau_1\tau_2}\right)\right).
\label{eq:probability}
\end{multline}
In our analysis, we acquired three data sets, consisting of approximately $82\times 10^3$ pairs of photon arrival times, for the three pump settings with different spectral width $\Delta \lambda$ specified in \tabref{tab:fitvalues}. For each data set we computed a histogram, to which we subsequently fitted the distribution given in \eqref{eq:probability}. We took into account the background noise in the model, which turned out to be negligible. The parameters were fitted using standard nonlinear model fitting functions. The best fit parameters, $\rho_t$, $\tau_1$ and $\tau_2$ are gathered in \tabref{tab:fitvalues}. 

\begin{figure*}[]
	\centering
	\includegraphics[width=2\columnwidth]{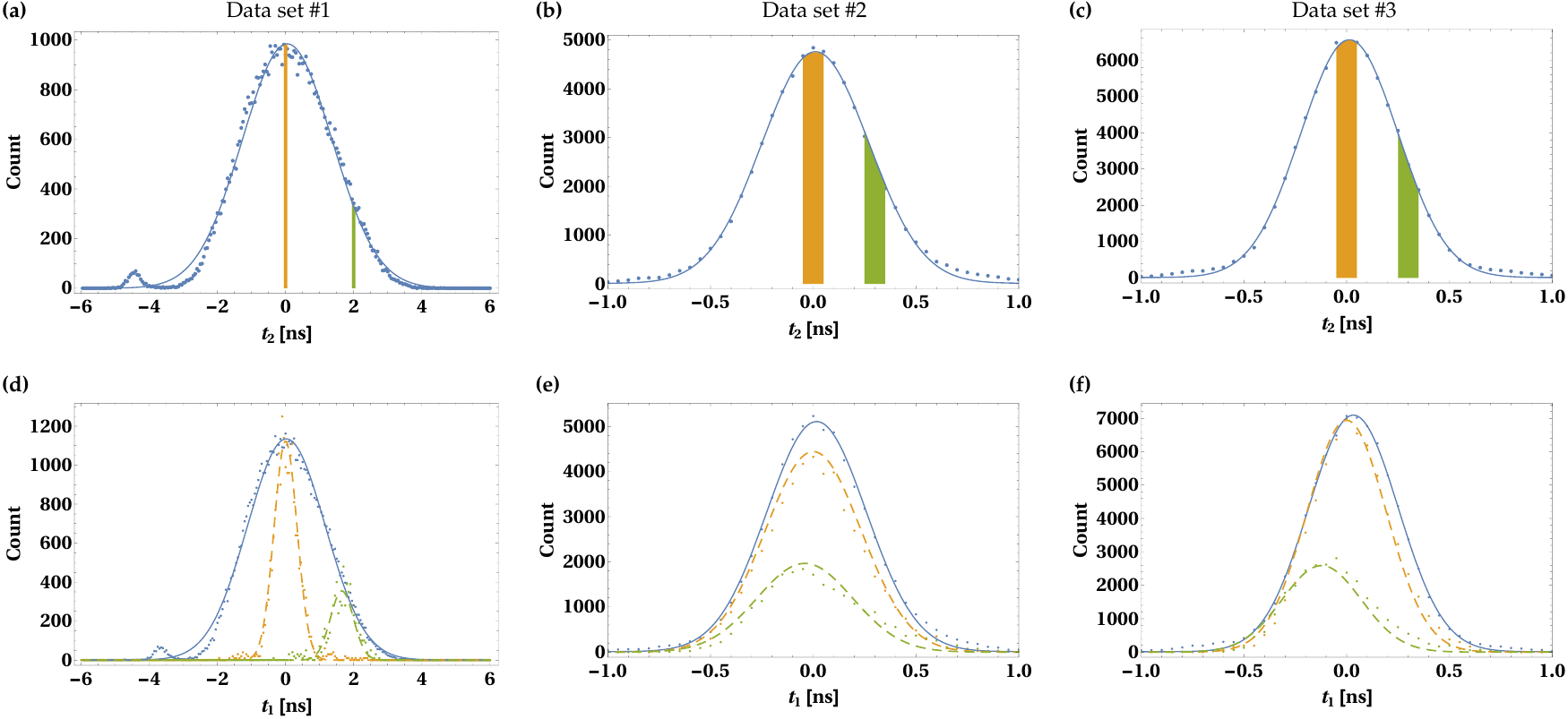}
	\caption{The top (bottom) row shows the distribution of arrival time of photon number two (one). Blue dots (lines) represent experimental data (fitted theoretical model given by \eqref{eq:p:con}). Each column corresponds to one of the three data sets specified in \tabref{tab:fitvalues}. The orange and green colored areas on panels (a) -- (c) denote arbitrary chosen, $100$ ps wide detection time windows for photon number two. The measured (calculated) distribution of photon number one arrival times corresponding to those selected areas are plotted on panels (d) -- (f) with dots (dashed lines) of the the same colors. The numbers of coincidences in the heralded photons peaks were scaled for convenience presentation of the wavepacket narrowing effect. On panels  (d) -- (f), the total number of coincidences in peaks of dashed yellow (green) plots are 125 (48), 865 (368) and 1396 (562).}
	\label{fig:wpnarrowing}
\end{figure*}

Let us now assume that the photon number two was detected at the time $T_2$ with uncertainty $\Delta T$, which is the width of detection window. In this case the probability distribution for the arrival time of the photon number one reads:
\begin{multline}
p_c\left(t_1;T_2,\Delta T\right)=\frac{\displaystyle\int_{T_2-\Delta T/2}^{T_2+\Delta T/2}\mathrm{d}t_2p(t_1,t_2)}{\displaystyle\int \mathrm{d}t_1\int_{T_2-\Delta T/2}^{T_2+\Delta T/2}\mathrm{d}t_2\,p(t_1,t_2)}=\\
\frac{e^{-\frac{t_1^2}{2 \tau_1^2}} \left(\text{erf}\left(\frac{ \tau_1\left(\Delta T-T_2\right)+\rho_t  \tau_2 t_1}{\sqrt{2-2 \rho_t ^2} \tau_1 \tau_2}\right)+\text{erf}\left(\frac{\tau_1 (\Delta T+T_2)-\rho_t  \tau_2 t_1}{\sqrt{2-2 \rho_t ^2} \tau_1 \tau_2}\right)\right)}{\sqrt{2 \pi } \tau_1 \left(\text{erf}\left(\frac{T_2+\Delta T}{\sqrt{2} \tau_2}\right)-\text{erf}\left(\frac{T_2-\Delta T}{\sqrt{2} \tau_2}\right)\right)}.
\label{eq:p:con}
\end{multline}
An analogous formula can be derived for the opposite situation, \emph{i.e.} when the photon number one heralds the photon number two. Note that $\Delta T \rightarrow \infty$ when the information on the arrival time of the heralding photon is not available for the experimenter.

The measured statistics are depicted in \figref{fig:wpnarrowing}, where the columns correspond to the three data sets specified in \tabref{tab:fitvalues}. The top (bottom) row refers to the photon number two (one). The blue solid lines denote theoretical prediction for the case of no timing information on the heralding photon. The profiles are given by the \eqref{eq:p:con} with the detection window set to $\Delta T \rightarrow \infty$. We also set two detection windows of width $\Delta T= 100$ ps centered on some randomly chosen values of the heralding time $T_2$. They are marked by orange and green stripes in the top row of \figref{fig:wpnarrowing}. The corresponding statistics of the heralded photon arrival times are depicted with respective colors in the bottom row. The dashed lines show the theoretical prediction.  

\begin{figure}[h]
	\centering
	\includegraphics[width=0.95\columnwidth]{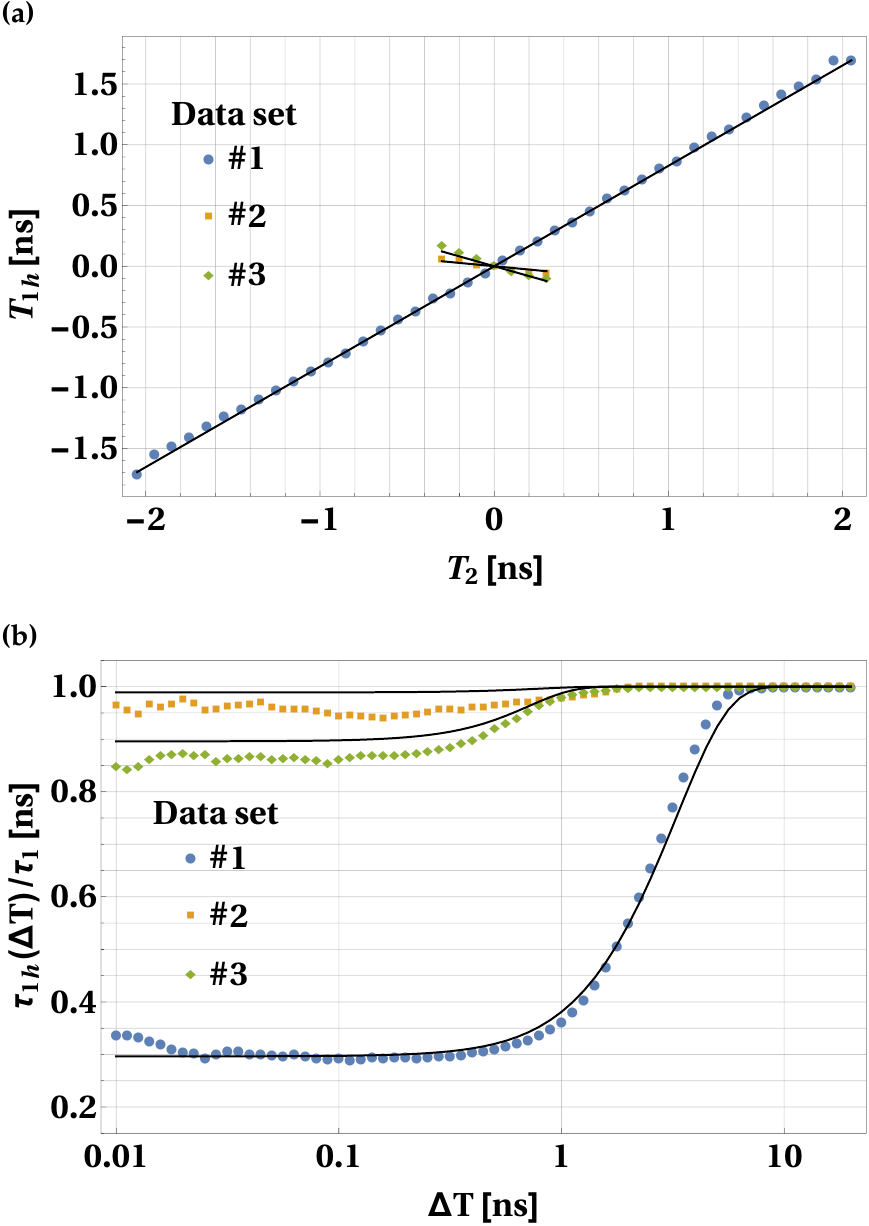}
	\caption{Panel (a) shows the ratio between the temporal widths of the heralded photon and non-heralded photon, plotted as a function of the width of the detection window for the heralding photon. Panel (b) shows the dependence of the position for the peak of the arrival time distribution function for the heralded photon, $T_{1h}$, on the position of the peak of the analogous function for the heralding photon, $T_2$. On both panels the blue, yellow and green dots correspond to the experimental data obtained for the three pump settings specified in the first, second and third column of the \tabref{tab:fitvalues}, respectively. The results of theoretical calculation are represented by black solid lines.}
	\label{fig:taut}
\end{figure}

For the finite detection windows $\Delta T$ we observe that the temporal width of the photon number one is reduced  as compared to the situation when there is no information on the heralding photon arrival time. The strength of this narrowing effect depends on the temporal correlation between the SPDC photons, which in turn depends on the pump settings. In general, the higher the absolute value of $\rho_t$, the lower the ratio of $\tau_{1h} (\Delta T)/\tau_1$. This can be seen in \figref{fig:taut}(a), where we plotted the experimental values of this ratio as a function of the width of the detection window, $\Delta T$. This data is also compared with the results of theoretical model depicted with solid lines. The value of $\tau_{1h} (\Delta T)/\tau_1$ is reduced when $\Delta T$ decreases. However, below a certain threshold value of $\Delta T$ it approaches a certain limit, which in theory is given by 
\begin{equation}
\lim\limits_{\Delta T \rightarrow 0 }\frac{\tau_{1h}(\Delta T)}{\tau_{1}}=\sqrt{1-\rho_t^2}.
\label{eq:ratio:limit}
\end{equation}
In this situation the conditional probability distribution for the detection time of the photon number one reads
\begin{multline}
\lim\limits_{\Delta T\rightarrow0}p_c\left(t_1;T_2,\Delta T\right)\!=\!\frac{1}{\sqrt{2\pi\left(1-\rho_t^2\right)}\tau_1}\\\times\!\exp\!\left(\frac{\left(T_2\rho_t\tau_1-t_1\tau_2\right)^2}{2(-1+\rho_t^2)\tau_1^2\tau_2^2}\right).
\label{eq:limpc}
\end{multline}
In our experiment, the aforementioned threshold value, below which further shortening of the detection window does not make the heralded wavepacket narrower, can be estimated to $\Delta T\approx300$ ps.
The strength of the wavepacket narrowing that we were able to observe, calculated in terms of the ratio of $\tau_{1h} (\Delta T)/\tau_1$, is given in \tabref{tab:fitvalues}.

The average arrival time of the heralded photon, $T_{1h}$, depends on the average detection time of the heralding one, $T_2$. The experimental results are depicted in  \figref{fig:taut} (b). In general there is no analytical formula to express this dependence. However, if one chooses detection window close to the limiting value $\Delta T\rightarrow 0$, the theoretical dependence is linear. It can be derived from \eqref{eq:limpc} and it reads:
\begin{equation}
T_{1h}=\rho\frac{\tau_1}{\tau_2}T_2
\label{eq:T:limit}
\end{equation}
In order to illustrate it, for every $T_2$ we took fit parameters from \tabref{tab:fitvalues} and numerically calculated $T_{1h}$. The results of this calculation are depicted by solid black lines in \figref{fig:taut} (b). 

\section{Optimizing the photon pair source}

\begin{figure}[t]
	\centering
	\includegraphics[width=0.95\columnwidth]{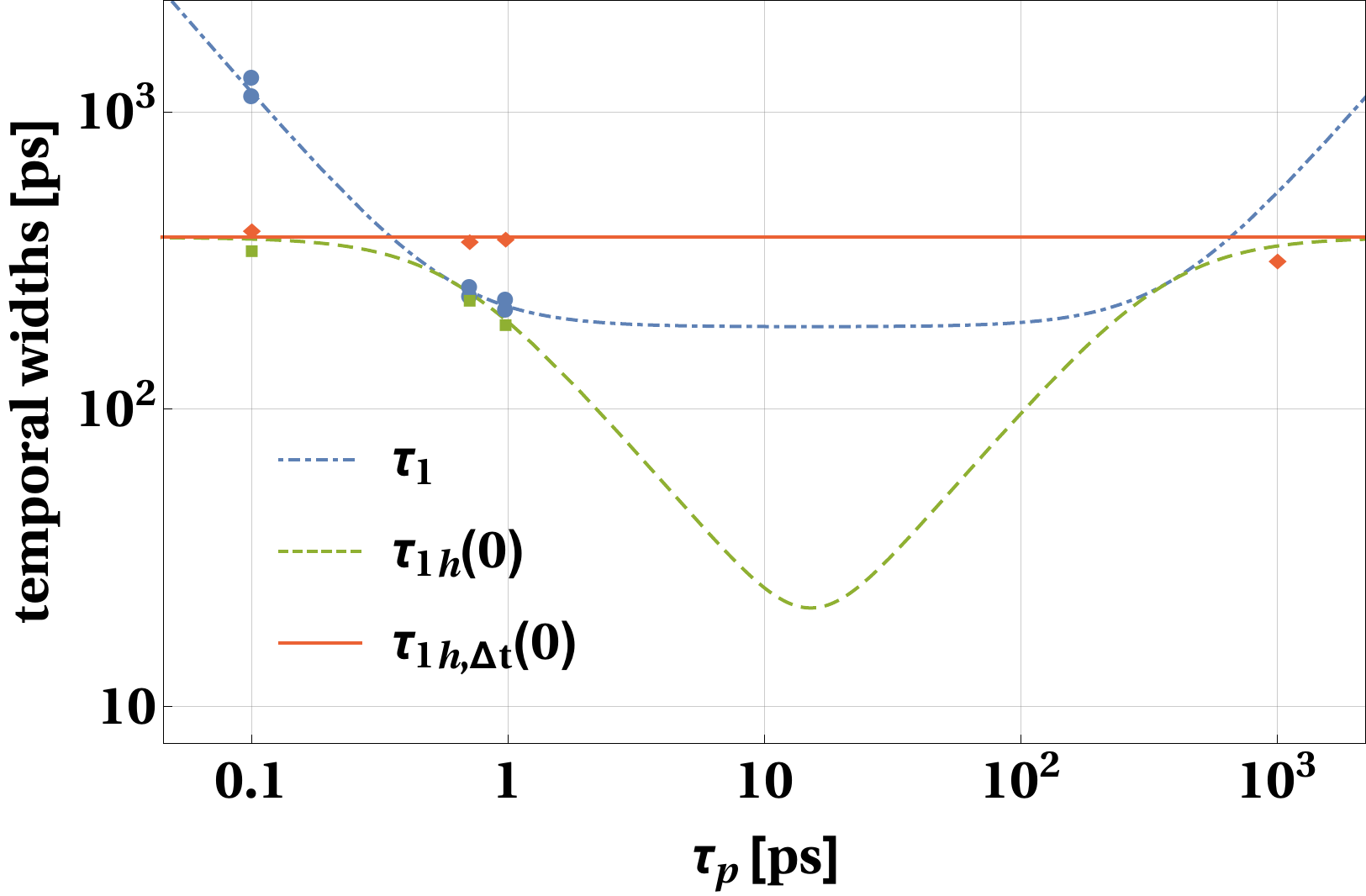}
	\caption{Temporal widths $\tau_1$ (blue, dot-dashed line), $\tau_{1h}(0)$ (green, dashed line) and $\tau_{1h,\Delta t}(0)$ (red, solid line) as functions of $\tau_p$, plotted for $\sigma=3.29$ THz, $L=10$ km and $\beta=-1.15\times10^{-26}$ s$^2/$m. Blue dots, green squares and red diamonds represent the values of $\tau_1$, $\tau_{1h}(0)$ and $\tau_{1h,\Delta t}(0)$ (respectively) obtained in the experiment. The rightmost green diamond corresponds to the experimental value of $\tau_{1h,\Delta t}(0)$ obtained for the CW pump laser.}
	\label{fig:taus10}
\end{figure}
\begin{figure*}[]
	\centering
	\includegraphics[width=2\columnwidth]{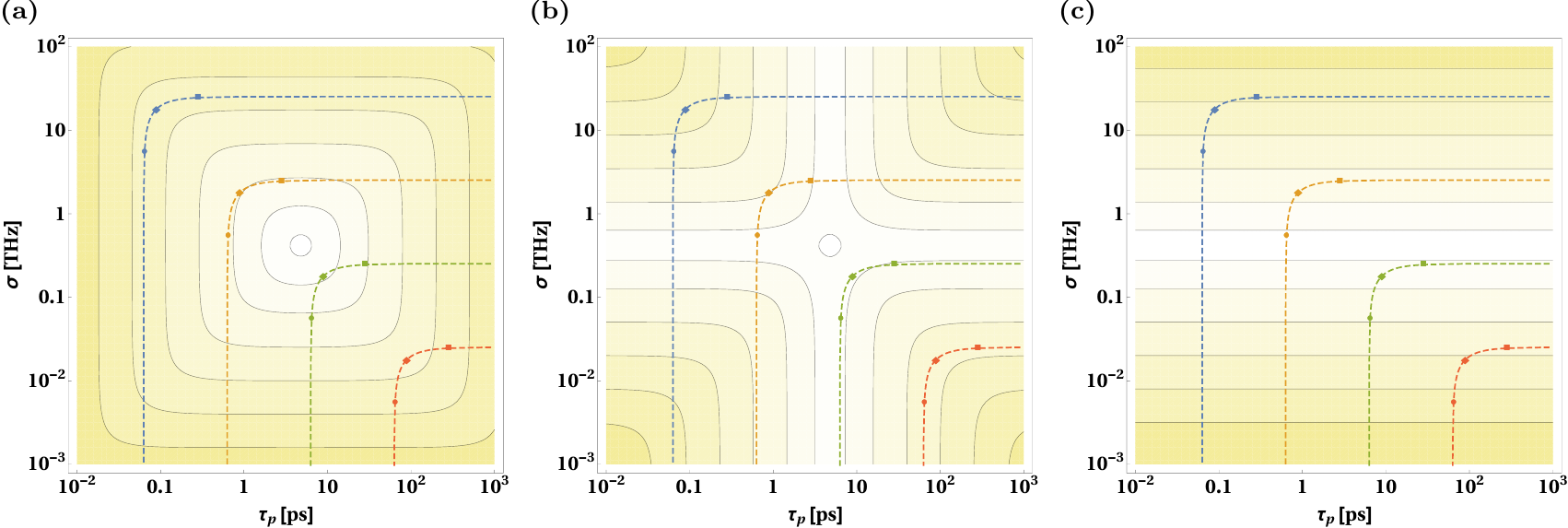}
	\caption{Logarithm of the temporal widths a) $\tau_1$, b) $\tau_{1h}(0)$ and c) $\tau_{1h,\Delta t}(0)$ plotted as functions of 
		$\sigma$ and 
		$\tau_p$. Dashed lines connect pairs of values $(\tau_p,\sigma)$ corresponding to the following spectral widths of photons emitted from the crystal: $\sigma_0=10\,\mathrm{GHz}$ -- red line (the closest one to the lower right corner of the pictures), $\sigma_0=100\,\mathrm{GHz}$ -- green line, $\sigma_0=1\,\mathrm{THz}$ -- orange line and $\sigma_0=10\,\mathrm{THz}$ -- blue line (the closest one to the upper left corner of the pictures). The colored dots, diamonds and squares denote pairs of points $(\tau_p,\sigma)$ for which the spectral correlation coefficient takes the value of $\rho=0.9$, $\rho=0$ and $\rho=-0.9$ (respectively). 
		The contours shown in each panel have the same value. The lines close to the darkest and  lightest filled regions correspond to value $-9.2$ and $-11.3$, respectively.}
	\label{fig:fulloptimization}
\end{figure*} 
The parametrization used in \eqref{eq:probability} allowed us to get compact forms of the expressions given by \eqref{eq:ratio:limit} and \eqref{eq:T:limit}, which are very useful in the description of the wavepacket narrowing. Now we wish to find the optimal source parameters that could provide us with the narrowest possible wavepackets for a given length of transmission links, $L$, characterized by the dispersion coefficient, $\beta$.
For this we assume a simplified biphoton wavefunction in the following form \cite{Lutz2014, Gajewski2016}:
\begin{equation}
\phi(\nu_1,\nu_2)=M\exp \left(-\frac{(\nu_1-\nu_2)^2}{\sigma^2}-\frac{(\nu_1+\nu_2)^2\tau_p^2}{4}\right),
\label{eq:Parametrization2}
\end{equation}
where $\sigma$ is the effective phase matching function width \cite{Kolenderski2009} and $\tau_p$ stands for the pump laser pulse duration.
This expression is equivalent to the formula (3) in Ref.~\cite{Sedziak2017} taken for equal spectral widths of the pair of photons produced in the crystal, \emph{i.e.} for $\sigma_1=\sigma_2\equiv\sigma_0$. 
The comparison between those two formulas gives the following transformation:
\begin{equation}
\begin{array}{lcl}
\tau_p\!\!\!\!\!\!& = &\!\!\!\! \frac{1}{\sigma_0\sqrt{1+\rho}},\\
\sigma\!\!\!\!\!\!& = &\!\!\!\! 2\sigma_0\sqrt{1-\rho}.
\end{array} 
\label{eq:transformation}
\end{equation}
This allows us to rewrite the formulas for the temporal widths of the heralded and non-heralded wavepackets derived in \cite{Sedziak2017} in terms of the parameters describing the photon pair source, $\sigma$ and $\tau_p$, and transmission links, $\beta$ and $L$.

Let us first focus on the temporal width of the non-heralded photon, which reads:
\begin{equation}
\tau_1=\sqrt{\frac{\sigma^2\tau_p^4+\left(\beta^2L^2\sigma^4+4\right)\tau_p^2+4\beta^2L^2\sigma^2}{4\sigma^2\tau_p^2}}.
\label{eq:tau1}
\end{equation}
It is not surprising that the value of $\tau_1$ grows to infinity when the pump pulse duration is infinitely short, $\tau_p\rightarrow 0$, or infinitely long, $\tau_p\rightarrow \infty$ (CW laser). 
On the other hand it has the minimum equal to
\begin{equation}
\tau_{1}^{\mathrm{min}}=\frac{|\beta| L\sigma^2+2}{2\sigma},
\label{eq:tau1opt}
\end{equation}
which is reached for $\tau_p^{\mathrm{opt}}=\sqrt{2|\beta| L}$.

In the asymptotic case, in which the arrival time of the heralding photon is perfectly known, one gets the following expression for the temporal width of the heralded photon:
\begin{equation}
\tau_{1h}(0)=\sqrt{\frac{\left(\beta^2L^2\sigma^4+4\right)\left(\tau_p^4+4\beta^2L^2\right)}{\sigma^2\tau_p^4+\left(\beta^2L^2\sigma^4+4\right)\tau_p^2+4\beta^2L^2\sigma^2}}.
\label{eq:tau1h}
\end{equation}
For $\tau_p^{\mathrm{opt}}$ this quantity also reaches its minimum, which is given by
\begin{equation}
\tau_{1h}^{\mathrm{min}}(0)=\frac{2\sqrt{|\beta| L\left(\beta^2L^2\sigma^4+4\right)}}{|\beta| L\sigma^2+2}.
\label{eq:tau1hopt}
\end{equation}

On the other hand, for the situation when the emission time of the pump pulse is unknown but the detection time of the heralding photon is available the temporal width of the heralded photon reads:
\begin{equation}
\tau_{1h,\Delta t}(0)=\frac{\sqrt{\beta^2L^2\sigma^4+4}}{\sigma}.
\end{equation}
It is interesting to note that for a given nonlinear crystal with fixed $\sigma$ parameter it does not depend on the pump settings. 


\subsection{Optimization over the pulse duration}
Let us first analyze the scenario where the crystal parameter, $\sigma$, is fixed and the experimenter can modify the pulse duration. 
The relation between $\tau_{1}$, $\tau_{1h}(0)$ and $\tau_{1h,\Delta t}(0)$ can be clearly seen in \figref{fig:taus10}, where the theoretically calculated and experimentally measured values of those functions are plotted. It can be seen that when the additional information about the photon pair generation time is available the respective temporal width is smaller compared to the case when there is no such information, $\tau_{1h}(0) < \tau_{1h,\Delta t}(0)$. This can be seen by comparing the solid red and the dashed green curves in \figref{fig:taus10}. 

Furthermore, one can observe that the temporal width of the heralded wavepacket is always not greater than non-heralded one, $\tau_{1h}(0)\leq\tau_1$. The strength of the narrowing effect, $\tau_{1h}(0)/\tau_1$,  asymptotically goes to zero for $\tau_p\rightarrow0$ and $\tau_p\rightarrow\infty$. On the other hand, for $\tau_p=2/\sigma$ and $\tau_p=|\beta|L\sigma$ it reaches the maximal value of one. The first maximum can be attributed to the crystal producing spectrally decorrelated photon pairs. The second one is the consequence of the propagation in the fiber. As it was shown in Ref.~\cite{Sedziak2017}, the temporal correlation changes with the propagation distance. Therefore the second maximum corresponds to the point where there is no temporal correlation. The local minimum of $\tau_{1h}(0)/\tau_1$ is reached for $\tau_p^{\mathrm{opt}}$. It can be calculated using the formulas (\ref{eq:tau1hopt}) and (\ref{eq:tau1opt}). For the case illustrated in Fig.\,\ref{fig:taus10}, this central minimum within our experimental setting takes the value of approximately $11.3\%$. It means that the lowest value of $\tau_{1h}(0)/\tau_1$ measured in our experiment, which reads $29.49\%$ (see \tabref{tab:fitvalues}), could be further reduced by setting the pulse duration to $\tau_p=15.2$ ps.

It is interesting to ask what type of spectral correlation do we have for the optimal pump settings, for which both $\tau_{1}$ and $\tau_{1h}(0)$ reach their minima. Through simple mathematical calculation one can derive the following formula for the optimal value of spectral correlation coefficient:
\begin{equation}
\rho^{opt}=\frac{2-|\beta| L\sigma^2}{2+|\beta| L\sigma^2}.
\end{equation} 
It means that for sufficiently short propagation distance the optimal SPDC photons are always positively correlated, while for very long distance communication schemes the best is negative correlation. However, the distance at which the type of optimal spectral correlation changes from positive to negative depends on the parameter $\sigma$, describing the properties of the nonlinear crystal.

\subsection{Designing the optimal SPDC source}
It is also important to know what is the optimal photon pair source design when one can arbitrarily choose both the crystal and the pump laser settings, $\tau_p$ and $\sigma$, for a given pair of symmetric transmission links characterized by the parameters $L$ and $\beta$. The dependence of the temporal widths $\tau_1$, $\tau_{1h}(0)$ and  $\tau_{1h,\Delta t}(0)$ on the pump laser settings, plotted for fixed values of $L$ and $\beta$, can be seen in \figref{fig:fulloptimization}. A simple calculation shows that both $\tau_1$ and $\tau_{1h}(0)$ reach their absolute minima for $\tau_p^{\mathrm{opt}}=\sqrt{2|\beta| L}$ and $\sigma^{\mathrm{opt}}=\sqrt{{2}/{|\beta| L}}$.  Those minima are identical and read:
\begin{eqnarray}
\tau_{1}^{\mathrm{abs}}=\tau_{1h}^{\mathrm{abs}}=\sqrt{2|\beta| L}.
\end{eqnarray}
This means that for the optimal SPDC source, the narrowing effect is not present. Therefore, spectrally decorrelated pairs are optimal for the quantum communication in this case. For our 10 km-long SMF fiber, links we get $\tau_p^{\mathrm{opt}}\approx15.2\,\mathrm{ps}$ and $\sigma^{\mathrm{opt}}\approx132\,\mathrm{GHz}$. On the other hand, the absolute minimal value of $\tau_{1h,\Delta t}(0)$ equals 
\begin{equation}
\tau^\text{abs}_{1h,\Delta t}(0)=2\sqrt{|\beta| L}.
\end{equation}
Again, it can be reached for $\sigma^{\mathrm{opt}}=\sqrt{{2}/{|\beta| L}}$, but for arbitrary $\tau_p$.


\section{Discussion}

The possibility of narrowing temporal widths of photons by performing the time-resolved measurements and optimizing the settings of the photon pair source, discussed above, can have multiple applications. As an example one can consider the problem of a distant clock synchronization. The experimental realization of this application presented by Valencia \textit{et al.} in Ref.~\cite{Valencia2004} was performed with SPDC source. However, the choice of the CW laser is certainly not optimal, as can be seen from Fig.\,\ref{fig:taus10}. Therefore, the results can be potentially improved by choosing a pulsed pump laser and optimizing its spectrum analogously as in our work. Similar optimization can also refine the results presented by Quan \textit{et~al.} in Ref.~\cite{Quan2016}, where the SPDC source used in quantum clock synchronization experiment was pumped by a pulsed laser, but no optimization of it was performed.  The results of our work can be also useful in extending the maximal secure distance of quantum communication schemes. This possibility was  discussed in Ref.~\cite{Sedziak2017}. Our new outcomes can  in principle further  improve the security analysis presented in that paper.

In summary, we investigated the problem of a wavepacket shaping of a single photon, which is heralded by the time-resolved detection of the other photon from SPDC pair. We showed how the strength of the wavepacket narrowing depends on the parameters of the photon pair source and the width of the detection window for the heralding photon. Our theoretical predictions \cite{Sedziak2017} were compared with the experimental results showing very good agreement. We experimentally observed the reduction of the width of the heralded wavepacket to approximately $29$\% as compared to the case of non-heralding scenario. Moreover, we performed the  optimization over the pump laser pulse duration in order to find the minimal possible wavepacket temporal width. For the case of our  experimental setup further narrowing down to $11$ \% is feasible for $\tau_p=15.2$ ps. Finally, we derived formulas for optimal parameters of the SPDC source minimizing the aforementioned temporal widths. In our experimental scenario both photons from SPDC pairs propagated through a pair of identical single-mode fibers. A generalization of this work to the case of fibers with non-equal lengths or different dispersion scenario is currently under our consideration. The results of our work can be applied to optimize the performance of various realizations of quantum communication protocols.

\section*{Funding Information}
Foundation for Polish Science (FNP) (project First Team co-financed by the European Union under the European Regional Development Fund); Ministry of Science and Higher Education, Poland (MNiSW) (grant no.~6576/IA/SP/2016; statutory R\&D activities supporting the development of young scientists and PhD students, internal grant no. 2821-F/2017); National Science Centre, Poland (NCN) (grant no.~2016/23/D/ST2/02064, grant no.~2016/23/N/ST2/02133), National Laboratory FAMO, Torun, Poland.



%

\end{document}